\documentclass[aps, prb, amsmath, twocolumn, 10pt, superscriptaddress]{revtex4-2}

\usepackage{bbold}
\usepackage{threeparttable}
\usepackage{booktabs}
\usepackage{graphicx}
\usepackage{braket}
\usepackage{upgreek}

\newcommand{\SiGe}{Si$_{1-y}$Ge$_y$}
\newcommand{\GeSn}{Ge$_{1-x}$Sn$_x$}
\newcommand{\h}[1]{\text{H}{#1}}
\newcommand{\e}[1]{\text{L}{#1}}

\usepackage{xcolor}

\begin{document}

\title{Fully Tunable Strong Spin-Orbit Interactions in Light Hole Germanium Quantum Channels}

\author{Patrick Del Vecchio}
\email{p.delvecchio@tudelft.nl}
\affiliation{Department of Engineering Physics, \'Ecole Polytechnique de Montr\'eal, Montr\'eal, C.P. 6079, Succ. Centre-Ville, Montr\'eal, Qu\'ebec, Canada H3C 3A7}
\affiliation{QuTech and Kavli Institute of Nanoscience, Delft University of Technology, Delft, Netherlands}
\author{Stefano Bosco}
\affiliation{QuTech and Kavli Institute of Nanoscience, Delft University of Technology, Delft, Netherlands}
\author{Daniel Loss}
\affiliation{Department of Physics, University of Basel, Klingelbergstrasse 82, 4056 Basel, Switzerland}
\author{Oussama Moutanabbir}
\email{oussama.moutanabbir@polymtl.ca}
\affiliation{Department of Engineering Physics, \'Ecole Polytechnique de Montr\'eal, Montr\'eal, C.P. 6079, Succ. Centre-Ville, Montr\'eal, Qu\'ebec, Canada H3C 3A7}

\begin{abstract}
 Spin-orbit interaction (SOI) is a fundamental component for electrically driven spin qubits and hybrid superconducting-semiconducting systems. In particular, Rashba SOI (RSOI) is a key mechanism enabling all-electrical spin manipulation schemes. However, in common planar systems, RSOI is weak because of the small mixing between heavy holes (HH) and light holes (LH), and instead relies on complex strain and interface phenomena that are hard to reliably harness in experiment. Here, MOS-like epitaxial Ge on relaxed \GeSn{} is introduced and shown to exhibit an inherently large, highly gate-tunable RSOI that is compatible with both spin qubits and hybrid devices. This large RSOI is a consequence of the LH-like ground state in Ge. Notably, the built-in asymmetry of the device causes the RSOI to completely vanish at specific gate fields, effectively acting as an on/off SOI switch. The LH $g$-tensor is less anisotropic than that of state-of-the-art HH qubits, alleviating precise magnetic field orientation requirements. The large in-plane $g$-factor also facilitates the integration of superconductors. Moreover, the out-of-plane $g$-factor is strongly gate-tunable and completely vanishes at specific gate fields. Thus, this material system combines the large RSOI with the scalability of planar devices, paving the way towards robust spin qubit applications and enabling access to new regimes of complex spin physics.
\end{abstract}

\maketitle

\footnotetext[1]{Details on the $k\cdot p$ model, the perturbative framework, the cubic Rashba parameters, the consistency of the quantum channel theory at large $l_x$ and the \GeSn{} material parametrization are provided in the Supplemental Material}

\paragraph*{}
The large spin-orbit interaction (SOI) intrinsic to semiconductor hole systems is a defining characteristic that provides the foundation for practical spin qubit devices. In fact, SOI enables all-electrical manipulation schemes such as electric dipole spin resonance (EDSR), positioning hole-spin qubits as one of the most promising candidates for scalable quantum processors~\cite{Bulaev2005PRL,Borsoi2024,Hsiao2024,Ivlev2024,Wang2023,Lawrie2023,vanRiggelen2021,Hendrickx2021,Hendrickx2020N,Lawrie2020APL,Hendrickx2020NC,Hardy2019}. In this context, Ge has emerged as a leading material for hole-spin qubits~\cite{Scappucci2020}, due to its large intrinsic SOI, absence of valley degeneracies, compatibility with Si processing, the inherently weak coupling of holes with surrounding nuclear spins~\cite{Philippopoulos2020,Testelin2009}, and the availability of nuclear spin-depleted Ge~\cite{Moutanabbir2024}. Therefore, the development of quantum device structures to control SOI in Ge has been an active area of research in the pursuit of reliable and robust qubits. 

The type of SOI is directly related to the dimensionality of the quantum device~\cite{Bihlmayer2015,Lyanda-Geller2022,Bosco2022,Bosco2021PRL}. For example, in quasi-one-dimensional systems, strong two-axis confinement leads to a large mixing of the heavy hole (HH) and light hole (LH) valence bands even at zero longitudinal momentum $k_z$, giving rise to the so-called direct Rashba SOI effect~\cite{Kloeffel2018,Luo2017,Kloeffel2011,Csontos2009}. This linear-in-$k_z$ SOI is highly tunable with transverse DC gate fields~\cite{Adelsberger2022_2,Adelsberger2022_1}, and plays a significant role in the electrical driving of the spin~\cite{Yu2023,Liu2023,Piot2022,Camenzind2022,Wang2022,Froning2021_Nat,Froning2021_PRR,Xu2020,Watzinger2018,Vukusic2018,Liu2018,Wang2016,Maurand2016,Higginbotham2014,Nadj-Perge2010}. Moreover, cold spots where SOI is turned off, at specific gate fields, have been reported~\cite{Bosco2021PRX,Kloeffel2013,Kloeffel2011}, which can be leveraged as a switch to change the qubit in an operational state or in an idle state. On the other hand, quasi-two-dimensional systems, such as the conventional planar Ge/\SiGe{} heterostructure~\cite{Lawrie2020APL}, do not show significant HH-LH mixing at zero momentum. As a result, electrical driving in these systems relies on a cubic-in-$k$ component of Rashba SOI (RSOI), in addition to contributions from strain-induced $g$-tensor tilts and wavefunction deformations~\cite{Hendrickx2024}. Notwithstanding the tunability with respect to the dimensions of the quantum well and electromagnetic fields~\cite{Wang2024,Wang2021,Terrazos2021,Bulaev2007,Bulaev2005PRB,Bulaev2005PRL} and enhancements by interface symmetry effects~\cite{Xiong2021,RodriguezMena2023} and inhomogeneous strain fields~\cite{Liles2021,RodriguezMena2023,AbadilloUriel2023}, the component of RSOI compatible with electrical driving is weak and remains a challenge to harness in experiments~\cite{Martinez2022}. These properties of conventional planar strained Ge/\SiGe{} heterostructures limit the prospects of this potentially scalable system.

\begin{figure}[th]
  \centering
  \includegraphics[width=\linewidth]{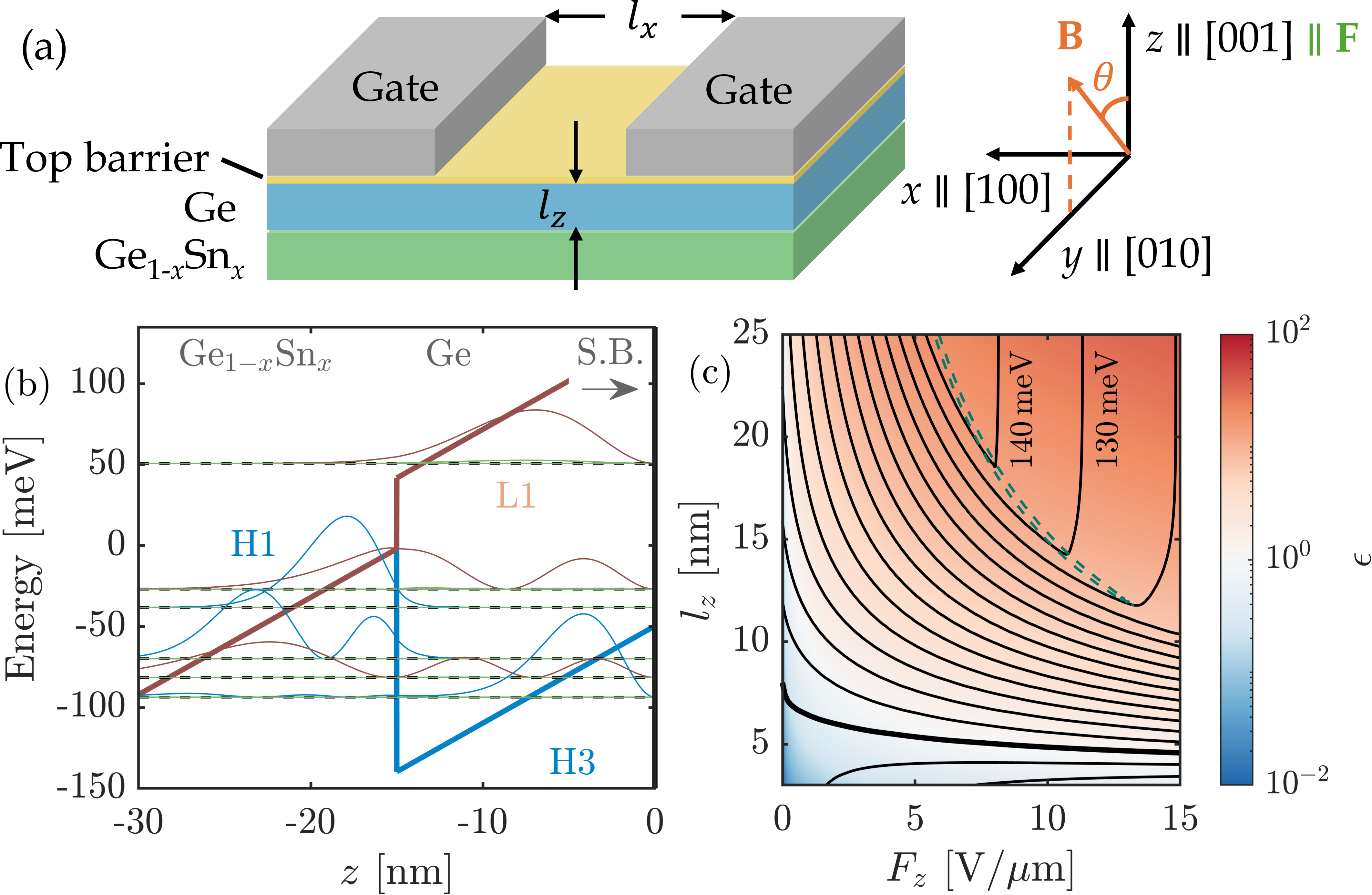}
  \caption{(a) Schematic of the Ge/\GeSn{}  quantum channel device with Cartesian axes and fields. (b) Energy confinement profile along the growth direction. Sn content in the relaxed barrier is $x=10\%$, the Ge well thickness is $l_z=15\,\text{nm}$, and the static electric field strength $F_z=6\,\text{V}/\upmu\text{m}$. The $z$-domain covers the \GeSn{} layer ($z<-l_z$) and the Ge layer ($-l_z<z<0$). The surface barrier (S.B.) is located at $z=0$. Blue envelopes correspond to the HH part of the hole spinor, while red and green envelopes correspond to the LH and the spin-orbit split-off part of the spinor, respectively. (c) Colormap of the ratio $\epsilon \equiv l_z^2/l_F^2$ as a function of $l_z$ and $F_z$, along with contour lines of the $\Delta E = E_1^\e{}-E_1^\h{}$ energy splitting ($10\,\text{meV}$-steps contour lines). Sn content is $x = 16\%$. The thicker black line indicates $\Delta E = 0\,\text{meV}$. $\h{1}$ and $\h{2}$ are $2\,\text{meV}$ apart on the dashed green line.}\label{fig:device}
\end{figure}

These limitations can be alleviated in  heterostructures with a LH ground state. To this end, recent progress in the growth of high-quality epitaxial \GeSn{} planar heterostructures offers this possibility, thus bringing another promising low-dimensional system for engineering RSOI. Tensile strained Ge quantum wells (QW) grown on relaxed \GeSn{} barriers can host LH ground states~\cite{Assali2022} with large linear-in-$k$ RSOI~\cite{DelVecchio2023,DelVecchio2024}. Consequently, spin-qubit devices based on scalable planar Ge/\GeSn{} heterostructures (Fig. \ref{fig:device}a) would benefit from large RSOI inherited directly from the planar geometry, rather than from transverse confinements or from hard-to-harness strain/interface effects. Herein, based on recent theoretical developments on LH spin physics~\cite{DelVecchio2023,DelVecchio2024}, LH spin dynamics in Ge/\GeSn{} gate-defined quantum channels (QC) and elongated quantum dots (QD) is investigated, revealing SOI that is strongly tunable by gate fields and compatible with electrical driving of spin qubits defined in quantum dots and  superconducting-semiconducting hybrid devices.

\paragraph*{Rashba SOI in planar systems.---} RSOI is a momentum-dependent $\mathbf{B}=0$ spin-splitting resulting from the spatial inversion asymmetry of the confining potential~\cite{Winkler2003}. In particular, an out-of-plane DC gate field applied perpendicularly to the plane introduces inversion asymmetry, giving rise to a predominantly cubic-in-$k$ spin splitting for HHs~\cite{Winkler2003,Moriya2014}. It is described by the effective Hamiltonian $H_{R3} = i\beta_2 k_-^3\sigma_+ - i\beta_3 k_+ k_- k_+\sigma_+ + \text{h.c.}$, where $\sigma_\pm = (\sigma_x \pm i\sigma_y)/2$ are the spin Pauli raising and lowering operators, $k_\pm = k_x\pm ik_y$, and $\beta_{2,3}$ are the two cubic-in-$k$ Rashba parameters. For HHs in Ge, the term relevant for EDSR, $\beta_3$, is $(\gamma_2+\gamma_3)/(\gamma_2-\gamma_3)$-times smaller than $\beta_2$ (with $\gamma_2=4.24$ and $\gamma_3=5.69$ being Ge Luttinger parameters~\cite{Note1}) since it arises only from small valence band anisotropies and from coupling with neighboring LH levels~\cite{Moriya2014,Winkler2003}. As discussed in the following, LH ground states possess an inherently large linear-in-$k$ RSOI term $i\beta_1k_-\sigma_+ + \text{h.c.}$ compatible with EDSR, as well as a $\beta_3$ parameter $(\gamma_2+\gamma_3)/(\gamma_2-\gamma_3)$-times {\it larger} than $\beta_2$.

\paragraph*{The proposed heterostructure.---} The new material system consists of a MOS-like Ge epitaxial layer (the quantum well) deposited on a strain-relaxed $[001]$-oriented \GeSn{} layer (the barrier). The out-of-plane confinement is provided by the valance band energy offsets between Ge and \GeSn{}. The lattice mismatch between Ge and relaxed \GeSn{} generates epitaxial biaxial strain in the Ge QW and lifts the degeneracy between LHs and HHs at the $\Gamma$ point. The ensuing strain tensor in Ge is diagonal, with components $\varepsilon_{xx} = \varepsilon_{yy} \equiv \varepsilon_\parallel = a_x/a_0-1$ and $\varepsilon_{zz} = -2\varepsilon_\parallel c_{12}/c_{11}$, where $a_0$ is the lattice constant of Ge and $a_x$ is that of relaxed \GeSn{}, and $c_{11}$, $c_{12}$ are elastic constants for Ge. Since $a_x>a_0$, the epitaxial strain in the QW is tensile and therefore pushes the LH band closer to the bandgap, enabling LH ground states in Ge. The surface barrier on top of Ge is assumed to be a thin, large band offset material such that LHs remain inside the QW when subjected to large electric fields directed towards the surface. For instance, this material could be a very thin Si capping (a few \aa ngstr\"oms), an oxide, or a combination thereof.

The LH-HH degeneracy at the $\Gamma$ point is also broken by spatial confinement due to the effective mass difference between LHs and HHs, where the latter are usually energetically favored  due to their larger out-of-plane mass~\cite{Winkler2003}. Therefore, a LH ground state in Ge is only attainable with large tensile strains that can compete with the confinement-induced energy splitting~\cite{DelVecchio2024}. The confinement energy profile of the device is displayed in Fig. \ref{fig:device}b, where a static gate electric field $\mathbf{F} = F_z\mathbf{e}_z$ pushes the holes towards the surface barrier. The peculiar profile of the HH band creates an additional triangular well at the interface between \GeSn{} and Ge. Depending on the well thickness $l_z$ and electric field strength $F_z$, the HH ground state ($\h{1}$) may be located either at the interface (as in Fig. \ref{fig:device}b) or at the surface barrier. In both cases, the wavefunction overlap and its spread across the Ge/\GeSn{} interface enables LH-HH mixing between the ground LH subband ($\e{1}$) and surrounding HHs.

\paragraph*{Light hole ground state requirements.---} Contour lines of the LH-HH energy splitting $\Delta E\equiv E_1^\e{}-E_1^\h{}$ computed from 6-band $k\cdot p$ theory~\cite{Note1} are shown in Fig. \ref{fig:device}c as a function of $l_z$ and $F_z$. Three regions are identified as follows: the first corresponds to a global HH ground state. It is located at small $l_z$, to the bottom of the bold contour line indicating $\Delta E=0\,\text{meV}$. As $l_z$ decreases, tensile strain is unable to compensate the increasingly strong confinement from the well, which eventually pushes $\h{1}$ to the ground state. The second region corresponds to a global LH ground state, with $\h{1}$ located at the Ge/\GeSn{} interface (as in Fig. \ref{fig:device}b). This region is located directly to the top of the $0\,\text{meV}$ contour line, where $\Delta E$ is strongly dependent on $l_z$ and/or $F_z$. The third region also corresponds to a LH ground state, but with $\h{1}$ located in Ge (in analogy to $\h{3}$ in Fig. \ref{fig:device}b). This region is identifiable by the mostly $l_z$-independent $\Delta E$, which is expected for electric field-dominated confinements. A sharp boundary between regions 2 and 3 coincides with $E_1^\h{} - E_2^\h{}\ll\Delta E$. This is indicated by the dashed green line, corresponding to $E_1^\h{} - E_2^\h{}=2\,\text{meV}$. The abruptness of the transition scales with the hybridization strength between HH levels located on either side of the triangular barrier at the Ge/\GeSn{} interface. Hybridization increases with lower Sn content in \GeSn{}, corresponding to a smaller triangular barrier.

The two primary contributions to the ground state energy $E_1^\e{}$ are the band offset at the Ge/\GeSn{} interface and the force $eF_z$ on L1 from the gate field. The former contribution scales roughly as $1/l_z^2$, whereas the latter scales as $1/l_F^2$, where $l_F$ is a characteristic electric length $l_F^3 \equiv \alpha_0\gamma_\perp^p/(eF_z)$, with $\alpha_0 = \hbar^2/(2m_0)$ and $\gamma_\perp^p$ being the inverse of the dimensionless out-of-plane effective mass of $\e{1}$ (the $p$ superscript stands for ``planar''). The colormap in Fig. \ref{fig:device}c shows the ratio of the two lengths, $\epsilon\equiv l_z^2/l_F^2$, indicating wether the ground state energy is dominated by the gate field ($\epsilon\gg 1$) or by the Ge/\GeSn{} band offset ($\epsilon\ll 1$).

\paragraph*{Planar ground state properties.---} Kinetic and spin properties of the ground state in the Ge/\GeSn{} planar system are readily quantified by computing an effective two-dimensional Hamiltonian for $\e{1}$ with all other subbands $\e{j}$ ($j\geq 2$) and $\h{l}$ ($l\geq1$) treated perturbatively. For static out-of-plane magnetic fields $\mathbf{B} = B\mathbf{e}_z$ and up to $k^3$, $\e{1}$ is described by (assuming $E_1^\e{} = 0$)

\begin{equation}\label{eff_e}
  \begin{split}
    H_\text{eff}^p &= \alpha_0\gamma_\parallel^pK_\parallel^2 + \frac{\alpha_0}{2l_B^2}g_\perp^p\sigma_z \\
    &+ \left[\left(i\beta_1^p K_- - i\beta_2^p K_+^3 + i\beta_3^pK_-K_+K_-\right)\sigma_+ + \text{h.c.}\right],
  \end{split}
\end{equation}

\noindent where $\mathbf{K} = \mathbf{k} + e\mathbf{A}/\hbar$ is the mechanical wavevector with $\mathbf{A} = Bx\mathbf{e}_y$ being the vector potential, $K_\pm = K_x\pm iK_y$, $K_\parallel^2 = \{K_-,K_+\}/2$, and $1/l_B^2 = [K_-,K_+]/2 = eB/\hbar$. The effective parameters are the following: $\gamma_\parallel^p$ is the in-plane effective mass, $g_\perp^p$ is the out-of-plane $g$-tensor component, and $\beta_i^p$ with $i=1,2,3$ are the aforementioned RSOI parameters. The Schrieffer-Wolff transformation employed to derive Eq.~\eqref{eff_e} provides exact formulas for these parameters in terms of the envelopes and energies at $k_x=k_y=0$ and $B=0$~\cite{DelVecchio2023,DelVecchio2024,Note1}.

\begin{figure}[t!]
  \centering
  \includegraphics[width=\linewidth]{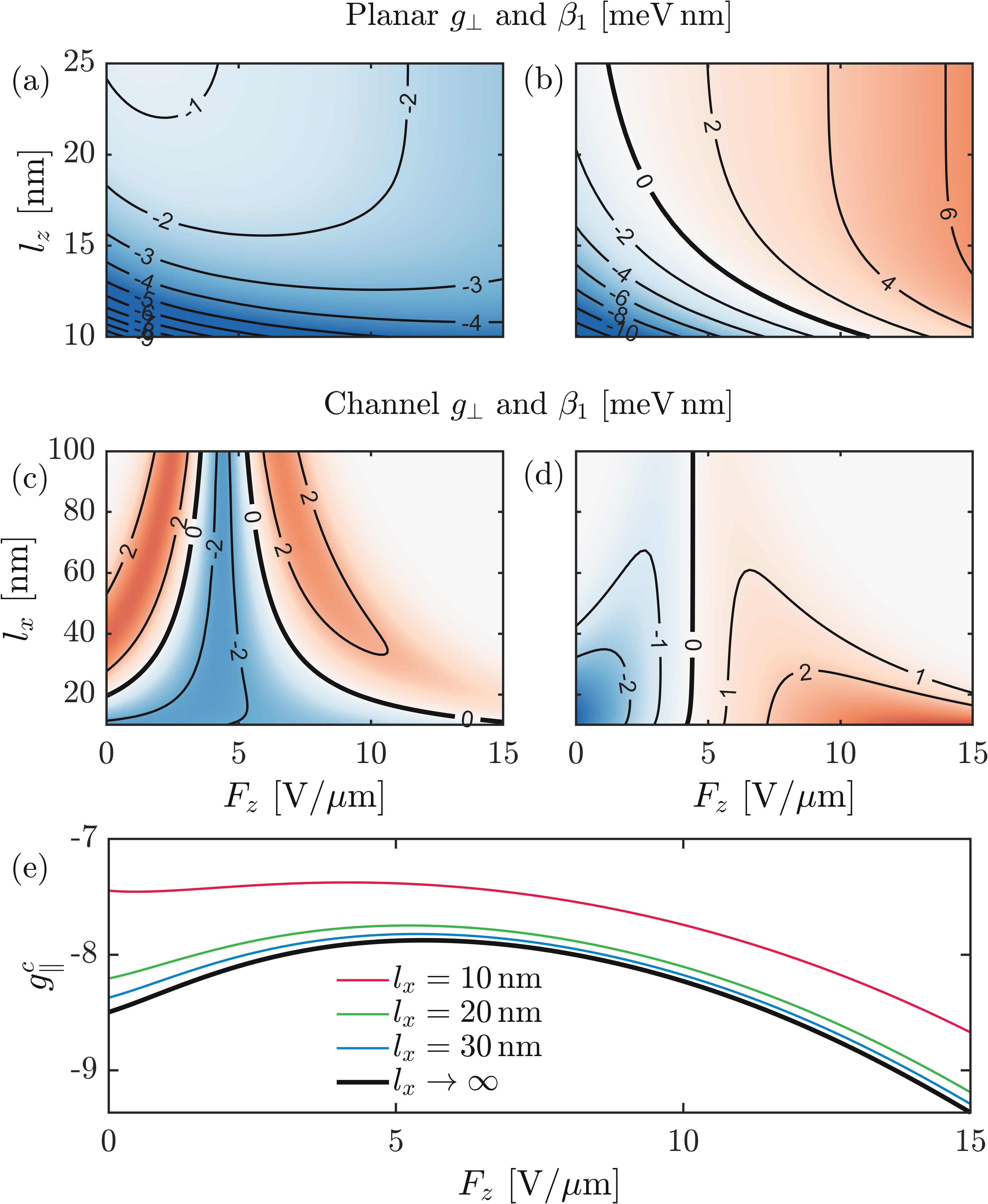}
  \caption{(a)-(b) $g_\perp^p$, and $\beta_1^p$ from Eq.~\eqref{eff_e} as a function of $F_z$ and $l_z$ with $x = 0.16$. The thick black line indicates $\beta_1^p = 0$. (c)-(d) $g_\perp^c$ and $\beta_1^c$ of the QC ground state as a function of $F_z$ and $l_x$ with fixed Sn content $x = 16\%$ and QW thickness $l_z = 15\,\text{nm}$. Colorcode in panels (a)-(d): blue (red) for negative (positive) values and white for zero. (e) $g_\parallel^c$ as a function of $F_z$ for selected $l_x$. The thick black line corresponds to the QW limit.}\label{fig:params}
\end{figure}

$g_\perp^p$ and $\beta_1^p$ are shown in Fig. \ref{fig:params}a-b as a function of $l_z$ and $F_z$ with Sn content $x=16\%$. $g_\perp^p$ slowly diverges for $l_z\lesssim 10\,\text{nm}$, which coincides with $\Delta E$ approaching $0\,\text{meV}$ and therefore leads to vanishing energy denominators in the perturbative expansion. In this regime, Hamiltonian \eqref{eff_e} fails to accurately describe $\e{1}$ because of strong couplings with neighboring subbands. On the other hand, $g_\perp^p$ and $\beta_1^p$ are well behaved for $l_z\gtrsim10\,\text{nm}$, and also become $l_z$-independent as $\epsilon\gg 1$. At a QW thickness $l_z=15\,\text{nm}$, $g_\perp^p\sim -3$ for the whole range of static electric fields. Most importantly, $\beta_1^p$ vanishes at finite $F_z$ due to the asymmetric confinement profile even at $F_z = 0$. The same is also true for the two other cubic Rashba parameters~\cite{Note1}. Notably, the $\beta^p=0$ contour line coincides in $(l_z,F_z)$-space for all three Rashba parameters, suggesting a fully gate-tunable RSOI.

\paragraph*{Quantum channel ground state properties.---} The SOI tunability with respect to $\mathbf{F}$ is now further investigated for a QC device. A parabolic in-plane confinement $V_\parallel = -\alpha_0 x^2/l_x^4$ is introduced towards the $x$-axis, such that the hole is free to move along the $y$-direction. Here, the characteristic length $l_x$ is interpreted as being relative to a hole sharing the same mass as a bare electron of mass $m_0$. The energy eigenstates in this geometry are denoted by $\ket{\tilde{\sigma};i,N}$ (with corresponding energy $E^c_{i,N}$), where $\tilde{\sigma}$ is the QC pseudo-spin index, $i\geq 1$ is an out-of-plane subband index and $N\geq 0$ is an in-plane orbital index. Spin properties of the QC ground state $\ket{\tilde{\sigma};1,0}$ are readily quantified by writing its two-dimensional effective Hamiltonian for $k_y\neq 0$ and $B>0$, with the coupling to all other levels $\ket{\tilde{\sigma};1,N\geq 1}$ and $\ket{\tilde{\sigma};i\geq 2,N\geq 0}$ treated perturbatively. The result is

\begin{equation}\label{eff_c}
  \begin{split}
    H_\text{eff}^c &= E_{1,0}^c + \alpha_0\gamma^ck_y^2 + \frac{\alpha_0}{2l_B^2}\left(g_\perp^c\cos\theta\tilde{\sigma}_3 + g_\parallel^c\sin\theta\tilde{\sigma}_2\right) \\
    &+ \left(\beta_1^c + \beta_3^ck_y^2\right)k_y\tilde{\sigma}_1.
  \end{split}
\end{equation}

\noindent The parameters in Eq.~\eqref{eff_c} are named similarly to those in \eqref{eff_e}, with a ``$c$'' superscript indicating the QC geometry. Hamiltonian \eqref{eff_c} is written within the gauge $\mathbf{A}(\mathbf{r}) = Bx\left(\cos\theta \mathbf{e}_y - \sin\theta \mathbf{e}_z\right)$, such that the field $\mathbf{B}$ lies within the $y$-$z$ plane with polar angle $\theta$. Hamiltonian \eqref{eff_c} contains a direct RSOI parameter $\beta_1^c = \hbar v_\text{so}$ which arises from the strong confinement along $z$ and the weaker confinement along $x$. It is analogous to the so-called spin-orbit velocity $v_\text{so}$ introduced in squeezed hole spin qubits~\cite{Bosco2021PRB}.

The QC effective parameters $g_{\perp,\parallel}^c$ and $\beta_1^c$ are displayed in Fig. \ref{fig:params}(c)--(d). A roughly $l_x$-independent $\beta_1^c = 0$ line is clearly visible in panel (d) near $F_z = F_z^* \approx 4.4\,\text{V}/\upmu\text{m}$, in agreement with $\beta_1^p = 0$ reported in panel (b) at the corresponding QW thickness. Panels (c) and (d) also reveal that near $F_z = F_z^*$ (or $\beta_1^p = 0$), $g_\perp^c\sim g_\perp^p$ for all $l_x$ considered. More generally, if $|\beta_1^p| l_x/\alpha_0\ll 1$, $g_\perp^c$ is well approximated by~\cite{Note1}

\begin{equation}\label{gpc}
  g_\perp^c \approx g_\perp^p + \frac{1}{\sqrt{\gamma_\parallel^p}}\left(\frac{\beta_1^p l_x}{\alpha_0}\right)^2,
\end{equation}

\noindent where $g_\perp^p < 0$. This is in close analogy to the exponential renormalization of the $g$-tensor reported in quasi-1D systems~\cite{Froning2021_PRR,Carballido2024}, but with an additional $g$-factor sign change for $l_x^2\geq\alpha_0^2 |g_\perp^p|\sqrt{\gamma_\parallel^p}/(\beta_1^p)^2$ (see Fig. \ref{fig:params}c). Compared to $g_\perp^c$, $g_\parallel^c\sim -8$ and remains finite for every $F_z$ and $l_x$ (see Fig. \ref{fig:params}e). The regime where $g_\perp^c\to 0$ and $\beta_1^c\to 0$ in panels (c)-(d) corresponds to the linear Rashba term dominating the in-plane confinement energy~\cite{Note1}. It is analogous to the decrease towards zero of the direct RSOI at very strong electric fields reported in Ge/Si core/shell nanowires and squeezed dots~\cite{Kloeffel2018,Bosco2021PRB}.

\begin{figure}[t]
  \centering
  \includegraphics[width=\linewidth]{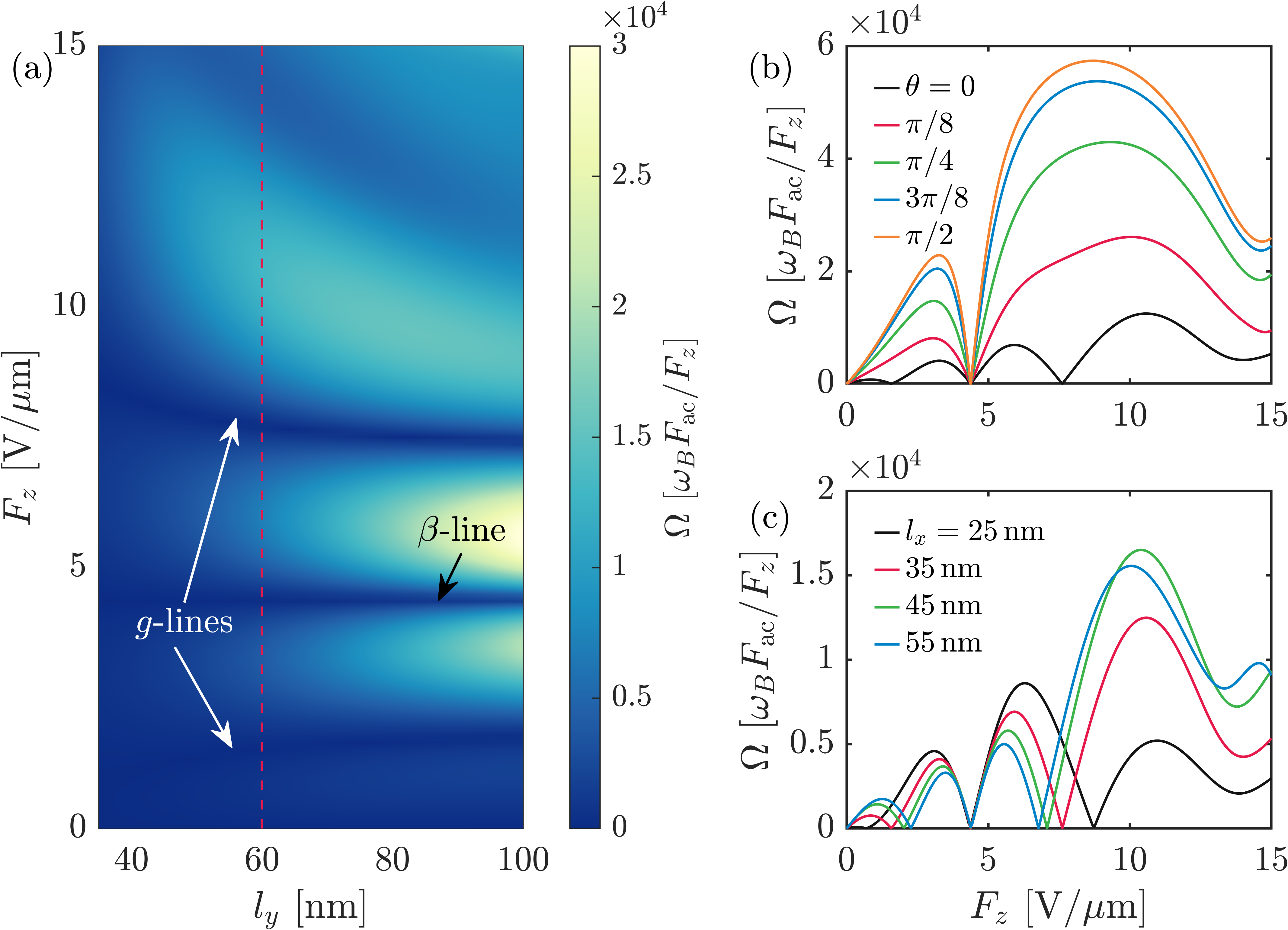}
  \caption{(a) Rabi frequency $\Omega$ as a function of $F_z$ and $l_y$, with $l_x=35\,\text{nm}$ and $\theta = 0$. (b) $\Omega$ along the dashed line in panel (a) for various $\theta$, with $l_x = 35\,\text{nm}$, $l_y = 60\,\text{nm}$ and $\phi = \pi/2$. (c) $\Omega$ for selected $l_x$, with $l_y = 60\,\text{nm}$ and $\theta = 0$.}\label{fig:rabi}
\end{figure}

\paragraph*{Quantum dot ground state properties.---} The effects of RSOI on the Rabi frequency of the LH ground state are investigated by introducing an additional parabolic confinement towards the $y$-direction, thus corresponding to a quantum dot (QD) system with characteristic lengths $l_x$ and $l_y\geq l_x$ in the $x$- and $y$-directions, respectively. An electrical driving $\tilde{\mathbf{F}} = F_\text{ac}\mathbf{e}_y\cos\omega t$ is applied along the axis of weak confinement, where $\hbar\omega$ is the QD ground state two-level system energy at finite $B$. The Rabi frequency $\Omega$ is estimated with $\Omega = (eF_\text{ac}/\hbar)\left|\bra{\mathbb{0}}y\ket{\mathbb{1}}\right|$, where $\ket{\mathbb{0}}$ and $\ket{\mathbb{1}}$ are the two-level system states. The magnetic field frequency $\omega_B \equiv \mu_\text{B}B_0/\hbar$ is also introduced, where $B_0 = 0.05\,\text{T}$ is the magnitude of the applied $\mathbf{B}$-field defining the qubit.

Fig. \ref{fig:rabi}a shows the computed Rabi frequency $\Omega$ in normalized units as a function of $l_y>l_x = 35\,\text{nm}$ and $F_z$ with a perpendicular-to-plane $\mathbf{B}$-field. An $\Omega = 0$ line at $F_z = F_z^*$ is clearly visible and corresponds to $\beta_1^c = 0$. This line is referred as the $\beta$-line. In addition, two more lines of $\Omega = 0$ appear near $F_z \approx 2\,\text{V}/\upmu\text{m}$ and $F_z \approx 7.5\,\text{V}/\upmu\text{m}$, and correspond to $g_\perp^c = 0$ with $l_x=35\,\text{nm}$ and $l_y\to\infty$. These lines are referred as the two $g$-lines. Fig. \ref{fig:rabi}b plots $\Omega$ along the dashed red line in panel (a) for various $\mathbf{B}$-field polar angles $\theta$. $\Omega$ increases as the field is tilted in the QW plane, in accordance with $g_\parallel^c>g_\perp^c$. Moreover, the two $g$-lines disappear as soon as $\mathbf{B}$ acquires an in-plane component, since the effective $g$ factor $g_\theta^c$ in the tilted configuration is $(g_\theta^c)^2 = (g_\perp^c\cos\theta)^2 + (g_\parallel^c\sin\theta)^2$, and $g_\parallel^c$ never vanishes (see Fig. \ref{fig:params}e). On the other hand, the $\beta$-line remains unperturbed for all $\theta$. A similar plot of $\Omega$ along the dashed red line is shown in panel (c) for $\theta = 0$ and for selected values of $l_x$. Here again, the $\beta$-line remains unperturbed by variations of $l_x$, whereas the two $g$-lines slightly shift towards the $\beta$-line with increasing $l_x$, in accordance with the $g_\perp^c = 0$ behavior shown in Fig. \ref{fig:params}c. The stability of the $\beta$-line with respect to $l_x$, $l_y$ and $\theta$ in the QD system is attributed to the linear RSOI being inherited directly from the planar geometry, i.e. $\beta_1 = 0$ occurs even without in-plane confinement. In contrast, the $g$-lines originate from the combined effects of in-plane confinement and $\beta_1^p\neq 0$ [see \eqref{gpc}]. The complete tunability of the out-of-plane $g$-factor suggests that the proposed heterostructure may provide an implementation of the recently proposed spinless spin qubits~\cite{RimbachRuss2024}, while the large in-plane g factor makes it appealing for superconducting-semiconducting hybrid systems.

\paragraph*{Conclusion.---} This work introduced a MOS-like Ge/\GeSn{} planar heterostructure as a new material system for LH spin qubits with fully gate-tunable RSOI. In the {\it on} state, RSOI allows for efficient EDSR driving of the spin through strong linear-in-$k$ and cubic-in-$k$ terms, whereas in the {\it off} state, the qubit is put in idle mode. In addition, fully gate-tunable out-of-plane $g$-factors  provide a scalable implementation of qubits that are free of leakage states if $g_\perp\approx0$~\cite{RimbachRuss2024}. On the other hand, the large in-plane $g$-tensor component of the LH spin alleviates precise magnetic field orientation requirements. This new heterostructure leverages the potential of LH spins as qubits for hybrid systems and for scalable quantum computing.

\paragraph*{Acknowledgments.---} We gratefully acknowledge the hospitality of the University of Basel, where part of this research was conducted. We acknowledge financial support
from NSERC Canada, Canada Research Chairs, Canada Foundation for Innovation, Mitacs, Mitacs Globalink under Award Number: FR106658, PRIMA Qu\'ebec, Defence Canada (Innovation for Defence Excellence and Security, IDEaS), the European Union’s Horizon Europe research and innovation program under Grant Agreement No 101070700 (MIRAQLS), the Air Force Office of Scientific and Research Grant No. FA9550-23-1-0763, NCCR Spin Grant No. 51NF40-180604 and the Army Research Office under Award Numbers: W911NF-23-1-0110 and W911NF-22-1-0277. The views and conclusions contained in this document are those of the authors and
should not be interpreted as representing the official policies, either expressed or implied, of the Army Research Office or the U.S. Government. The U.S. Government is authorized to reproduce and distribute reprints for Government purposes notwithstanding any copyright notation herein.

%

\end{document}